\documentclass[letterpaper]{JHEP3}

\usepackage{graphicx}

\usepackage{amsmath}
\usepackage{amsfonts}
\usepackage{epsfig}

\newcommand{\beq}{\begin{equation}}
\newcommand{\eeq}{\end{equation}}

\newcommand{\cF}{{\cal F}}

\newcommand{\cD}{{\cal D}}
\newcommand{\cDb}{{\overline{\cal D}}}
\newcommand{\cQ}{{\cal Q}}
\newcommand{\cU}{{\cal U}}
\newcommand{\cUb}{{\overline{\cal U}}} 

\newcommand{\Tr}{{\rm Tr\;}}
\newcommand{\bx}{{\bf x}}
\newcommand{\bmu}{{\boldsymbol \mu}}
\newcommand{\bnu}{{\boldsymbol \nu}}

\newcommand{\be}{{\boldsymbol e}}

\newcommand{\half}{\frac{1}{2}}

\title{First results from simulations of supersymmetric lattices}

\author{Simon Catterall,
Department of Physics, Syracuse University, Syracuse, NY 13244, USA
E-mail: \email{smc@phy.syr.edu}
}

\preprint{}

\date{October 2008}

\abstract{
We conduct the first numerical simulations of lattice theories with
exact supersymmetry arising from the orbifold constructions of
\cite{Cohen:2003xe,Cohen:2003qw,Kaplan:2005ta}. 
We consider the
$\cQ=4$ theory in $D=0,2$ dimensions and the
$\cQ=16$ theory in $D=0,2,4$ dimensions.
We show that the $U(N)$ theories do not possess
vacua which are stable non-perturbatively, 
but that this problem can be circumvented after truncation
to $SU(N)$. We measure the distribution of scalar field eigenvalues, the
spectrum of the fermion operator and the phase of the
Pfaffian arising after integration over the fermions. We monitor supersymmetry
breaking effects by measuring a simple Ward identity. Our results
indicate that simulations of ${\cal N}=4$ super Yang-Mills  
may be achievable in the near future.
}


\begin{document}

%
\section{Introduction}

The study of supersymmetric theories on lattices has a long history --
see the 
review \cite{Feo_rev1} 
and references
therein. Recently there has been a resurgence of interest in the
field with successful constructions of lattice theories which keep
intact a subalgebra of
the full supersymmetry algebra \cite{Kaplan:2003uh,Catterall:2005eh,Giedt_rev1}.

In this paper we will be concerned with specific discretizations
of the $\cQ=4$ and $\cQ=16$ supercharge
Yang-Mills
theories in a variety of dimensions. The lattice actions we employ were
first derived using orbifold/deconstruction techniques in 
\cite{Cohen:2003xe,Cohen:2003qw,Kaplan:2005ta}.
Recently, it was shown how to recover them by discretization of a twisted
version
of the target super Yang-Mills theories \cite{Catterall:2007kn}\footnote{The connection
between topological twisting and orbifold constructions had been anticipated
earlier in \cite{Damgaard:2007xi,Unsal:2006qp,Takimi:2007nn} and has since been
generalized in \cite{Damgaard:2008pa}}. Other proposals for lattice
actions based on twisting can be found in 
\cite{Sugino_sym1,Sugino_2d,Sugino_sym2,D'Adda_2d,D'Adda_super,Kato_bf}.
We will use the language of the twisted constructions in this paper.

The key feature of these actions is that, in addition to
gauge invariance, they retain one or more exact
supersymmetries at non vanishing lattice spacing. Since this feature is
already sufficient to pair each bosonic state with a fermionic state
of the same energy it is expected that these models may flow to the
target continuum theory with a minimum of fine tuning as the lattice
spacing is sent to zero.

In this paper,
we show that the vacua of these lattice theories with
$U(N)$ gauge symmetry exhibit a non-perturbative
instability associated to a runaway trace mode of the scalar
fields. We show that this problem can be evaded if the gauge group
is truncated to $SU(N)$ at the price of a mild breaking of the exact
supersymmetry. In two dimensions and for the $\cQ=16$ theory
we show that, nevertheless, supersymmetry is restored
without fine tuning in the continuum limit. The situation in the
two-dimensional 
$\cQ=4$ theory is more problematic on account of the
observed strong fluctuations
in the phase of the
Pfaffian that results from the fermion integration. These fluctuations
lead to large statistical
errors in all observables making it difficult to draw definite conclusions.

In contrast we observe rather small phase fluctuations for the
$\cQ=16$ theory with gauge group $SU(2)$ in both two
and four dimensions for the
small lattices used in this study.

The organization of the paper is as follows;
we first summarize the lattice actions with which we are concerned, describe
some of the details of the simulation algorithms that
are employed and then summarize
our numerical results for the case of $\cQ=4$ and $\cQ=16$ supercharge
theories in dimensions zero, two and four. We end with a discussion 
and outlook for the future.

\section{Lattice actions}

\subsection{$\cQ=4$ theory in two dimensions}

The field content of the lattice theory comprises a multiplet of
p-form fermions $(\eta,\psi_\mu,\chi_{12})$ distributed over sites
and links of the lattice together with a complexified Wilson
gauge link $\cU_\mu$. These fields transform under a 
scalar supersymmetry as follows:
\begin{eqnarray}
\cQ\; \cU_\mu&=&\psi_\mu\nonumber\\
\cQ\; \psi_\mu&=&0\nonumber\\
\cQ\; \cUb_\mu&=&0\nonumber\\
\cQ\; \chi_{\mu\nu}&=&\cF^{\dagger}_{\mu\nu}\nonumber\\
\cQ\; \eta&=&d\nonumber\\
\cQ\; d&=&0
\end{eqnarray}
Furthermore, as was shown in \cite{Catterall:2007kn} the 
action of the theory can be written in a $\cQ$-exact form 
\beq
S=\kappa\cQ\left[\sum_{\bx}\Tr\left( \chi_{\mu\nu}\cF_{\mu\nu}+\eta 
\cDb^{(-)}_\mu \cU_\mu-\frac{1}{2}\eta d\right)\right]
\label{qexact4}\eeq
where $\kappa$ is a dimensionless bare coupling.
The decomposition of the fermions into p-forms and the appearance
of a scalar supersymmetry arises as a consequence of
the twisting procedure which was described in detail
in \cite{Catterall_n=2,Catterall_n=4}. 

Notice that this supersymmetry is nilpotent making the supersymmetric
invariance of the the lattice action in eqn.~\ref{qexact4} manifest.
In this formulation the gauge links are {\it non-unitary} matrices of
the form $U_\mu(x)=e^{A_\mu(x)+iB_\mu(x)}$ with $\mu=1,2$. 
The imaginary parts of the
gauge fields generate the scalar fields of theory in the continuum limit
as was shown in \cite{Catterall:2007kn}
\footnote{We use an antihermitian basis for the generators}.
The gauge covariant difference operators appearing in eqn.~\ref{qexact4}
are defined by
\begin{eqnarray}
\cD^{(+)}_\mu f_\nu(\bx)&=&\cU_\mu(\bx)f_\nu(\bx+\be_\mu)-
f_\nu(\bx)\cU_\mu(\bx+\be_\nu)\\
\cDb^{(-)}_\mu f_\mu(\bx)&=&f_\mu(\bx)\cUb_\mu(\bx)-
\cUb_\mu(\bx-\be_\mu)f_\mu(\bx-\be_\mu)
\label{derivs}
\end{eqnarray}
where the unit lattice vectors are $\be_1=(1,0)$, $\be_2=(0,1)$
and the lattice field strength $\cF_{\mu\nu}$ 
is given by
\beq
\cF_{\mu\nu}=\cD^{(+)}_\mu
\cU_\nu(\bx)=\cU_\mu(\bx)\cU_\nu(\bx+\bmu)-
\cU_\nu(\bx)\cU_\mu(\bx+\bnu)\label{field}\eeq 
In general the lattice fields
are associated to links of the lattice and they transform correspondingly
under gauge transformations: 
\begin{eqnarray}
\eta(\bx)&\to&G(\bx)\eta(\bx)G^\dagger(\bx)\nonumber\\
\psi_\mu(\bx)&\to&G(\bx)\psi_\mu(\bx)G^\dagger(\bx+\be_\mu)\nonumber\\
\chi_{\mu\nu}(\bx)&\to&G(\bx+\be_\mu+\be_\nu)\chi_{\mu\nu}(\bx)G^\dagger(\bx)\nonumber\\
\cU_\mu(\bx)&\to&G(\bx)\cU_\mu(\bx)G^\dagger(\bx+\be_\mu)\nonumber\\
\cUb_\mu(\bx)&\to&G(\bx+\be_\mu)\cUb_\mu(\bx)G^\dagger(\bx)
\end{eqnarray}
Notice that this choice of link and
orientation for the twisted lattice fields maps exactly
into their r-charge assignments in the orbifolding approach
\cite{Cohen:2003xe}. Indeed, the lattice actions derived bv the two
methods are identical.

The final lattice action takes the form
\beq
S=\sum_{\bx}\Tr \left(\cF^{\dagger}_{\mu\nu}\cF_{\mu\nu}+
\frac{1}{2}\left(\cDb^{(-)}_\mu \cU_\mu\right)^2-
\chi_{\mu\nu}\cD^{(+)}_{\left[\mu\right.}\psi_{\left.\nu\right]}-
\eta \cDb^{(-)}_\mu\psi_\mu\right)
\label{4action}
\eeq
where we have integrated out the auxiliary field $d$.

It is also possible to consider dimensional reductions of this model
to lower dimensions. For example we will show results obtained from
simulations of the $D=0$ matrix model that results from this lattice theory
after dropping all
dependence on lattice coordinates $\bx$. The dimensionless lattice coupling
$\kappa$ in reduced dimension $D$ is given by
\beq\kappa=\frac{N}{2\lambda}\left(\frac{L}{\beta}\right)^{(4-D)}
\label{coupling}\eeq
where $\lambda=g^2N$ is the 
't Hooft coupling, $\beta$ is the physical extent of the system and $L$ the
lattice length.

\subsection{$\cQ=16$ theory in four dimensions}

Remarkably, the $\cQ=16$ supercharge theory in four dimensions also possesses
a $\cQ$-exact term in its action of precisely the same form 
as its two dimensional cousin:
\beq
S=\kappa\cQ\left[\sum_{\bx}\Tr\left( \chi_{ab}\cF_{ab}+\eta 
\cDb^{(-)}_a \cU_a-\frac{1}{2}\eta d\right)\right]
\label{qexact16}\eeq
where the indices now run $a,b=1\ldots 5$ and the sixteen fermions
of ${\cal N}=4$ super Yang-Mills
are now built from one scalar, five vectors and the ten components
of the antisymmetric tensor $\chi_{ab}$. The real parts of
$U_\mu,\mu=1\ldots 4$ yield the usual four dimensional gauge field
while $U_5$ and the remaining imaginary components of $U_\mu$ yield
the expected six scalar fields of ${\cal N}=4$ super Yang-Mills
\cite{Catterall:2007kn}.

The action of the nilpotent, scalar supersymmetry is the same as
before
\begin{eqnarray}
\cQ\; \cU_a&=&\psi_a\nonumber\\
\cQ\; \cUb_a&=&0\nonumber\\
\cQ\; \psi_a&=&0\nonumber\\
\cQ\; \chi_{ab}&=&\cF_{ab}^\dagger\nonumber\\
\cQ\; \eta&=&d\nonumber\\
\cQ\; d&=&0
\label{latticeQ}
\end{eqnarray}

In addition a new $\cQ$-closed term is needed of the
form 
\beq
S_{\rm closed}=-\frac{\kappa}{8}\sum_{\bx}\Tr 
\epsilon_{abcde}\chi_{de}(\bx+\be_a+\be_b+\be_c)
\cD^{(-)}_c\chi(\bx+\be_c)\label{closed}\eeq
which is zero by virtue of an exact Bianchi identity satisfied by the
lattice field strength
\beq
\epsilon_{abcde}D^{(+)}_c\cF_{de}=0\eeq
Gauge invariance of this term requires $\sum_a^5 \be_a=0$. 
This condition is satisfied if the basis vectors
are taken
to be $\be_a^i=\delta^i_j$ with $a=1\dots 4$ and $\be_5=(-1,-1,-1,-1)$
corresponding to their r-charge assignments in the orbifold
construction \cite{Kaplan:2005ta}. 
 
Again, the link assignments of the fields can be summarized 
by specifying the variation of the fields
under gauge transformations  
\begin{eqnarray}
\eta(\bx)&\to& G(\bx)\eta(\bx) G^\dagger(\bx)\nonumber\\
\psi_a(\bx)&\to& G(\bx)\psi_a(\bx) G(\bx+\be_a)\nonumber\\
\chi_{ab}(\bx)&\to&G(\bx+\be_a+\be_b)\chi_{ab}(\bx)G^\dagger(\bx)\nonumber\\
\cU_a(\bx)&\to&G(\bx)\cU_\mu(\bx)G^\dagger(\bx+\be_a)\nonumber\\
\cUb_a(x)&\to&G(\bx+\be_a)\cUb_\mu(\bx)G^\dagger(\bx)\\
\end{eqnarray}
The action of the covariant difference operators appearing
in the action take the same form as for the $\cQ=4$ theory eqn.~\ref{derivs}.
In the naive continuum limit it is possible to show that this
lattice theory reduces to the Marcus topological twist 
of ${\cal N}=4$ super Yang-Mills \cite{Marcus} (in modern parlance the GL-twist
\cite{gl})
and hence in flat space is fully equivalent to the usual 
continuum theory.

Again, the model can be reduced to lower dimensions by simply taking the
fields to be independent of certain lattice coordinates and we will
show data for the model in both zero, two and four dimensions. The
bare coupling $\kappa$ is again given by eqn.~\ref{coupling}.

\section{Simulation details}

Since both the $\cQ=4$ and $\cQ=16$ supercharge theories have such a similar
structure they can be simulated using a single Monte Carlo code in which
the lattice geometry, gauge group and number of supercharges are
input as parameters.
The dynamical twisted fermions are handled exactly and efficiently
using the RHMC algorithm \cite{rhmc}. For completeness we
summarize the main features of this algorithm as applied to
the simulation of supersymmetric lattices here.

If we denote the set of
twisted fermions by the field $\Psi=(\eta,\psi_\mu,\chi_{\mu\nu})$ 
we first introduce a parallel pseudofermion field $\Phi$ with action
\beq
S_{\rm PF}=\Phi^\dagger 
(M^\dagger M)^{-\frac{1}{4}} \Phi\label{pseudo}\eeq
where $M=M(\cU,\cU^\dagger)$ is the antisymmetric twisted lattice fermion operator
given, for example, in eqn.~\ref{4action}\footnote{The antisymmetry
is guaranteed if the fermion action is rewritten as the sum
of the original terms plus their lattice transposes}. 

Integrating over the
fields $\Phi$ will then yield (up to a possible phase) the Pfaffian of
the operator $M(\cU,\cU^\dagger)$ as required. The fractional
power is approximated by the partial fraction expansion
\beq
\frac{1}{(M^\dagger M)^{\frac{1}{4}}}=\alpha_0+\sum_{i=1}^P\frac{\alpha_i}{M^\dagger
M+\beta_i}\label{partial}\eeq
where the coefficients $\{\alpha_i,\beta_i\}$ are evaluated offline using
the Remez algorithm to minimise the error in some interval $(\epsilon,A)$.
Typically we have used $P=15$ which yields a fractional error of
$0.00001$ for the interval $0.0000001\to 1000.0$ which conservatively
covers the range we are interested in.

Following the standard procedure we introduce momenta $(p_\cU,p_F)$ conjugate
to the coordinates $(\cU,\Phi)$ and evolve the coupled system using
a discrete time leapfrog algorithm according to the
classical Hamiltonian $H=S_B+S_{\rm
PF}+p_\cU\bar{p}_\cU+p_\Phi\bar{p}_\Phi$.
One step of the discrete time update is given by
\begin{eqnarray}
\delta p_\cU&=&\frac{\delta t}{2}\bar{f}_\cU\\
\delta p_\Phi&=&\frac{\delta t}{2}\bar{f}_\Phi\\
\delta\cU&=&\left(e^{\delta t p_\cU}-I\right)\cU\\
\delta\Phi&=&\delta t p_{\Phi}\\
\delta p_\cU&=&\frac{\delta t}{2}\bar{f}_\cU\\
\delta p_\Phi&=&\frac{\delta t}{2}\bar{f}_\Phi
\end{eqnarray}
where the forces $f_\cU$ and $f_\Phi$ are given by
\begin{eqnarray}
f_\cU&=&-\frac{\delta S}{\delta \cU}\\
f_\Phi&=&-\frac{\delta S}{\delta \Phi}
\end{eqnarray}
and the bar denotes complex conjugation. Using the partial fraction
expansion given in eqn.~\ref{partial} these forces take the form
\begin{eqnarray}
f_\cU&=&\sum_{i=1}^P\alpha_i\left[
\bar{t}_i\frac{\delta M}{\delta\cU}s_i+
\overline{\left(\bar{t}_i\frac{\delta M}{\delta\cUb}s_i\right)}
\right]\\
f_\Phi&=&-\alpha_0\bar{\Phi}-\sum_{i=1}^P\alpha_i\bar{s}_i\\
\end{eqnarray}
where 
\begin{eqnarray}
(M^\dagger M+\beta_i)s_i&=&\Phi\\
t_i&=&Ms_i
\end{eqnarray}
The latter set of sparse linear equations is solved using a multimass CG-solver
\cite{Jegerlehner:1996pm} which allows for the simultaneous solution of all $P$
systems in a single CG solve.
 
At the end of one such classical trajectory the final configuration
is subjected to a standard Metropolis test based on the Hamiltonian $H$.
The symplectic and reversible nature of the discrete time update
is then sufficient to allow for detailed balance to be satisfied
and hence expectation values are independent of $\delta t$.
After each such trajectory the momenta are refreshed from the
appropriate gaussian
distribution as determined by $H$ which renders the simulation
ergodic.

In the work reported here we have employed only periodic boundary
conditions which preserve supersymmetry. Thermal boundary conditions
are also interesting as they allow for exploration of
dualities between string and gauge theory 
\cite{Catterall:2007me,Catterall:2007fp,Catterall:2008yz,Kawahara:2007ib,
Hanada:2007ti,
Anagnostopoulos:2007fw}
and dynamical supersymmetry breaking \cite{Kanamori:2007yx,Kanamori:2008bk}.

Our measurements concentrate, in part, on local observables
such as the eigenvalues of $\cU^\dagger_\mu(x)\cU_\mu(x)$ and the
bosonic action $<S_B(\cU)>$. The former yields, in the continuum limit,
the distribution of eigenvalues of the scalar
fields $\cU^\dagger_\mu\cU_\mu-I=2B_\mu+\ldots$ 
and hence gives information on the quantum moduli
space. The latter observable is related to a exact supersymmetric
Ward identity and can be evaluated analytically which provides both a
useful check on our code and measures the magnitude of supersymmetry breaking
effects. The analytic argument that determines the value of
$<S_B>$ is simple. Consider first
the $\cQ=4$ supercharge theory and write down an expression for
the mean action $<S>$
\beq <S>=-\frac{\partial \ln{Z}}{\partial \kappa}=<\cQ\Lambda>=0\eeq 
where the last result follows from $\cQ$-exact nature of the twisted
action and shows that the vanishing mean
action is the consequence of a simple $\cQ$-Ward identity. This
argument needs a minor modification for the $\cQ=16$ supercharge theory
which contains also a $\cQ$-closed term. However, it is straightforward to
show that a simple rescaling of the field
$\chi_{\mu\nu}\to\sqrt{\kappa}\chi_{\mu\nu}$ removes the
$\kappa$ dependence of this term so that once again
$\frac{\partial \ln{Z}}{\partial\kappa}=0$
as a consequence of $\cQ$-supersymmetry\footnote{Another way to 
see this is to realize that the partition function for
periodic boundary conditions is just the Witten index and hence
does not depend on the coupling constant}. 

The result $<S>=<S_B>+<S_F>=0$ can be translated into an exact result for
the bosonic action since the fermions appear quadratically in the action
and hence $<S_F>$ can be evaluated by a simple scaling argument. In the
case of $\cQ=4$ supercharges one finds
\beq
\kappa<S_B>=\frac{3}{2}N_GV
\label{ward4}\eeq
where $N_G$ is the number of generators of the group and $V$ is the number
of lattice points.
One might have naively expected a factor of $4$ representing
the four twisted fermions of the $\cQ=4$ theory rather than the 
factor of $3$ that is present 
in this expression - the discrepancy arises as a consequence of
integrating out the auxiliary field $d$ which effectively removes
the contribution of one fermion to $<S_B>$.

The $\cQ=16$ theory is a little more involved. To evaluate $<S_F>$ by
a scaling argument requires an additional rescaling of both 
$\eta$ and $\psi_\mu$ by
a factor of $\sqrt{\kappa}$. This results in an additional multiplicative
factor of $\kappa^{6N_GV/2}$ in the measure. The final result
for $<S_B>$ is then
\beq
\kappa<S_B>=\frac{9}{2}N_GV
\label{ward16}\eeq
where the
factor arising in the numerator is now composed from $9=16-6-1$.

A faster way to derive these results relies simply on the 
coupling constant independence of the free energy;
the bosonic action can then be evaluated in the weak
coupling limit where the theory is quadratic in the bosons
and equipartition holds;
the bosonic action then simply
counts the number of degrees of freedom. 

For the small systems we have examined in this paper we have
also measured the
Pfaffian and the spectrum of the fermion operator
$M(\cU,\cUb)$. The Pfaffian computation is carried out by using a variant of
Gaussian elimination with full pivoting to transform the $2n\times 2n$
dimensional antisymmetric matrix $M$ into
the canonical form
\beq
\left(\begin{array}{ccccc}0 &\lambda_1&0 &0&\ldots \\
                         -\lambda_1 & 0 &0 &0&\ldots \\
			  0 & 0 & 0 &\lambda_2&\ldots\\
			  0 & 0 & -\lambda_2 & 0&\ldots\\
			  0 & 0 & 0 & 0 &\ldots\\
      \end{array}\right)\eeq
Then 
\beq {\rm Pf}(M)=\prod_{i=1}^n \lambda_i\eeq
As can be seen by examining eqn.~\ref{pseudo} our simulations generate
the {\it phase quenched} ensemble defined by $|Pf(M)|$.
As usual we can always
compensate for neglecting any phase $e^{i\alpha(\cU)}$
by re-weighting all observables $O(\cU)$ by
the phase factor according to the simple rule
\beq
<O>=\frac{<O(\cU)e^{i\alpha(\cU)}>_{\rm \alpha=0}}{<e^{i\alpha(\cU)}>_{\rm
\alpha=0}}\eeq
Where necessary we have reweighted our results accordingly.

\section{Zero dimensions}

\subsection{Vacuum instability for $U(N)$ theories}
The construction we have described is strictly valid only for
$U(N)$ theories. This can be seen in a variety of ways; taking the trace
of the first line of the $\cQ$-variations in eqn.~\ref{latticeQ} is
inconsistent if the fields are restricted to the traceless generators.
Similarly the difference operators employed in the
lattice action when applied to a traceless field
generically yield a field with non-zero trace. 
Thus we initially focus on simulations of the $U(N)$ theories. Unfortunately
we will see that these theories have a non-perturbative instability;
the trace mode of the scalar fields rapidly
runs off to (negative) infinity. 

\begin{figure}
\begin{center}
\includegraphics[height=60mm]{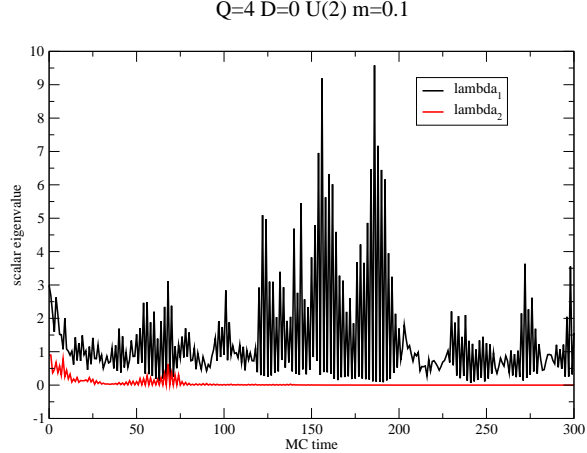}
\caption{Eigenvalues of $\cU^\dagger_\mu\cU_\mu$ versus
MC time. Gauge group $SU(2)$, $\cQ=4$ and $D=0$}
\label{figure1}
\end{center}
\end{figure}

Figure~\ref{figure1}
shows this explicitly with a plot of the Monte Carlo history of
the eigenvalues of $\cU^\dagger_\mu(x)\cU_\mu(x)$ for an $SU(2)$ model
with $\cQ=4$ supersymmetries reduced to zero dimensions. Clearly
one of the eigenvalues
is driven to zero after a finite simulation time. Using the
representation 
\beq\cU_\mu(x)=e^{A_\mu(x)+iB_\mu(x)}\label{exp}\eeq
it is easy to see
that ${\rm det}(M(x))=e^{2\sqrt{N}B^0_\mu(x)}$
where $B^0_\mu(x)$ is the trace component of the scalar field\footnote{Clearly one
zero eigenvalue is the minimum required for consistency with a
vanishing determinant -- we have also encountered runs where some
or all of the $\cU_\mu^\dagger\cU_\mu$ matrices have more than
one zero eigenvalue}. The
presence of a zero eigenvalue leads to a vanishing determinant and 
clearly is associated with a limit in which
the trace mode $B^0_\mu\to -\infty$. In this situation
the fluctuations in the gauge links are never small and no limit
exists in which the lattice model approximates the correct continuum
theory. Furthermore, we have observed that no simple scalar mass term
as advocated in \cite{Cohen:2003xe} is able to save the situation. 
Table~\ref{table1}
shows the value of $<\frac{1}{N}\Tr \cU^\dagger_\mu(x)\cU_\mu(x)>$ 
obtained from a series of runs in which the lattice action is
supplemented with the additional term
\beq
\Delta S=m^2\sum_x \left(\cU^\dagger_\mu(x)\cU_\mu(x)-I\right)^2\eeq

\begin{table}
\begin{center}
\begin{tabular}{||c|c|c|c||}\hline
$m$ & $S_B$ & $S_B^{\rm exact}$ & $\half\Tr(\cU^\dagger_\mu\cU_\mu)$\\\hline
0.01 & 2.65(6) & 6.0 & 0.45(2)\\\hline
0.1 & 2.40(8) & 6.0 & 0.57(6)\\\hline
0.5 & 2.99(7) & 6.0 & 0.38(2)\\\hline
\end{tabular}
\caption{Mass dependence of $\half\Tr\cU^\dagger_\mu\cU_\mu$}
\label{table1}
\end{center}
\end{table}
The first column gives the mean bosonic action as compared, in the
second column, to its exact value as predicted by eqns.~\ref{ward4} and
\ref{ward16}. Clearly, the $\cU_\mu$ fields deviate strongly
from the unit matrix for all values of the mass parameter $m$
and furthermore supersymmetry is also badly broken. In all these
cases we also observe a zero eigenvalue in $M$ independent of the mass. These
conclusions survive for other gauge groups $U(N)$, values of the
supercharge $\cQ$
and dimension of the model. 

It is straightforward
to see why this occurs; the bosonic
action evaluated on an arbitrary $\{\cU_\mu(x)\}$ configuration is
generically lowered by
a global shift in the trace mode $B^0_\mu(x)\to B^0_\mu(x)-c$ with positive
$c$\footnote{For a classical vacuum state corresponding to constant diagonal
matrices the bosonic action remains zero under this shift}. Explicitly,
\beq
S_B(\cU_\mu(x)e^{-cI})=e^{-4c}S_B(\cU_\mu(x))\eeq
since $e^{-cI}$ commutes with all terms in the bosonic action (each of which
contains a product of four $\cU_\mu$ matrices).
Unlike in the
continuum the penalty from the mass term is finite even in the limit
$B^0_\mu\to-\infty$ and entropic effects appear to drive the system
far from the point $B^0_\mu=0$. 
Notice that this effect occurs only for the exponential
parametrization of the complex 
$\cU_\mu$ matrices used here and given in eqn.~\ref{exp}. The
prescription utilized in the original orbifold constructions used
instead the decomposition
\beq
\cU_\mu=I+A_\mu+iB_\mu\eeq
which is not subject to this same instability.

One way to evade these problems immediately presents itself; simply
set the trace mode to zero by truncating the lattice model
to the special unitary group $SU(N)$. 
Notice that this truncation preserves supersymmetry
if the (complexified) gauge fields lie infinitessimally
close to the identity
$\cU_\mu(x)=I+aA_\mu(x)+\ldots$ which should happen in the continuum limit.
Presumably the restriction to $SU(N)$ is also
irrelevant in the large $N$ limit. However,
for non-zero lattice spacing and finite $N$ it will lead to supersymmetry
breaking effects 
which we examine in the next section.

\subsection{Vacuum structure for $SU(N)$ theories}

\subsubsection{$\cQ=4$ supercharges}

Once the truncation to $SU(N)$ is carried out we observe no instability;
the eigenvalues of $\cU^\dagger_\mu\cU_\mu$ cluster around unity
for any gauge group, dimension or number of supercharges. 
Furthermore the width of this eigenvalue distribution
decreases as the bare lattice coupling increases which is a necessary
condition for 
the lattice model to approximate the target
theory in the continuum limit.

Nevertheless, after this
truncation we might expect to see supersymmetry breaking effects and
we have probed for this by examining the bosonic action. As we have argued
earlier the expectation value of this term can be derived exactly
as a consequence of the twisted scalar supersymmetry. Deviations
from the exact value hence measure supersymmetry
breaking. Table~\ref{table2}
shows results from a simulation of the $\cQ=4$ model with $SU(2)$ gauge
group in the matrix model limit\footnote{The Pfaffian is real
positive definite for all $N$ in this limit and hence no
issue of reweighting occurs}.
\begin{table}
\begin{center}
\begin{tabular}{||c|c|c||}\hline
$\kappa$ & $<S_B>$ & $S_B^{\rm exact}$\\\hline
1.0 & 4.40(2) & 4.5 \\\hline
10.0 & 4.47(2) & 4.5 \\\hline
100.0 & 4.483(15) & 4.5 \\\hline
\end{tabular}
\caption{Bosonic action for $SU(2)$ $\cQ=4$ model at several couplings}
\label{table2}
\end{center}
\end{table}
While a small, statistically significant deviation is seen for strong
coupling this disappears for larger $\kappa$. The limit
$\kappa\to\infty$ corresponds to the continuum limit
$a\to 0$ for theories in $1\le D < 4$.
A similar pattern 
seen in table~\ref{table3} for the same model with group $SU(3)$.
\begin{table}
\begin{center}
\begin{tabular}{||c|c|c||}\hline
$\kappa$ & $<S_B>$ & $S_B^{\rm exact}$\\\hline
1.0 & 11.71(2) & 12.0 \\\hline
10.0 & 11.98(3) & 12.0\\\hline
100.0 & 11.98(4) & 12.0 \\\hline
\end{tabular}
\caption{Bosonic action for $SU(3)$ $\cQ=4$ model at several couplings}
\label{table3}
\end{center}
\end{table}

\begin{figure}
\vspace{1cm}
\begin{center}
\includegraphics[height=60mm]{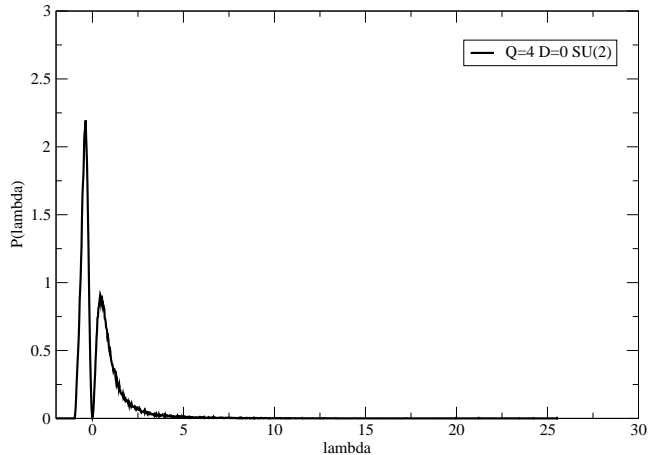}
\caption{Probability distribution of
eigenvalues of $\cU^\dagger_\mu\cU_\mu-I$ for $\cQ=4$ and $SU(2)$ and $D=0$}
\label{figure2}
\end{center}
\end{figure}

Having verified that the breaking of supersymmetry is indeed small
after the truncation from $U(N)$ to $SU(N)$
we now turn to the scalar field eigenvalues.
Figure~\ref{figure2} shows a plot of the probability distribution of
the eigenvalues of $\cU^\dagger_\mu\cU_\mu-I$ for $SU(2)$ gauge group.
Classically the model contains flat directions corresponding to
constant diagonal $\cU$ matrices. One might worry that the presence of
these flat directions might render the path integral defining the
quantum theory ill defined but these numerical results, like earlier
matrix model studies \cite{Krauth:1998xh,Krauth:1999qw}, confirm 
that the scalar fields
are localised around the origin in moduli space and the partition 
function is finite\footnote{As has been seen before \cite{Catterall_sims} the
distribution of scalar field eigenvalues is observed to
possess $N$ peaks for
group $SU(N)$}. The asymmetry in the distribution is a cut-off effect;
the lattice theory, unlike the continuum theory, is not invariant under a 
change of sign of the scalar fields. Indeed, the entire
region $-\infty < B_\mu < 0$ is mapped onto the finite segment
$-1< U^\dagger_\mu U_\mu-1 <0$.
The asymmetry in the distribution can be 
hence be used to quantity the magnitude of lattice
artifacts. Since the dimensionless eigenvalues contain a factor of
the lattice spacing the tails of the
distribution contract and are less sensitive to the hard cut-off
$\lambda=-1$ as the
lattice spacing is reduced. 

\begin{figure}
\begin{center}
\includegraphics[height=60mm]{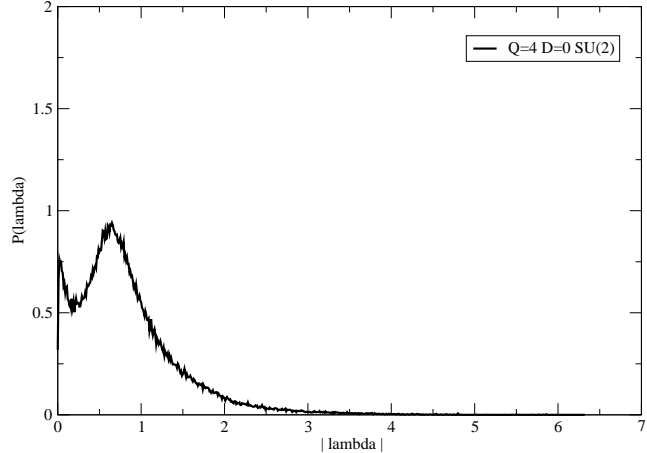}
\caption{Probability distribution of
eigenvalues of the fermion operator for $\cQ=4$, $SU(2)$ and $D=0$ }
\label{figure3}
\end{center}
\end{figure}
 
More importantly, notice that the effect of the classical flat
directions is still visible in the presence of a power law
tail in the distribution extending out to large (positive)
eigenvalue. Both
theoretical arguments and numerical results suggest the power $p=3$
independent of $N$ for this model \cite{Krauth:1999qw}. Such a power
behavior would yield a 
logarithmically divergent value for
$<\lambda^2>$ which means that in any Monte Carlo simulation the scalars
spend an appreciable amount of time far the origin in field
space. In the background of one of these field configurations the
bosonic action will necessarily develop a near zero mode corresponding
to translations along this flat direction. By supersymmetry we expect
the fermion operator must also then develop a near zero mode. This can
be seen explicitly in figure~\ref{figure3} which shows the distribution
of the absolute
value of the fermion eigenvalue for this system. An enhancement at small
eigenvalue is seen consistent with the previous argument.

\subsubsection{$\cQ=16$ supercharges}

In tables~\ref{table4},~\ref{table5} we examine
the bosonic action for
the $\cQ=16$ supercharge model in the matrix model limit.
\begin{table}
\begin{center}
\begin{tabular}{||c|c|c||}\hline
$\kappa$ & $<S_B>$ & $S_B^{\rm exact}$\\\hline
1.0 & 13.67(4)& 13.5 \\\hline
10.0 & 13.52 & 13.5\\\hline
100.0 & 13.48(2)& 13.5 \\\hline
\end{tabular}
\caption{Bosonic action for $SU(2)$ $\cQ=16$, $D=0$ model at several couplings}
\label{table4}
\end{center}
\end{table}
\begin{table}
\begin{center}
\begin{tabular}{||c|c|c|c||}\hline
$\kappa$ & $<S_B>$ & $S_B^{\rm exact}$& $<\cos{\alpha}>$\\\hline
1.0 & 36.01(20) & 36.0 & 0.67(5)\\\hline
10.0 & 35.95(18) & 36.0 & 0.47(12)\\\hline
100.0 & 35.92(10) & 36.0 & 0.47(13)\\\hline
\end{tabular}
\caption{Bosonic action for $SU(3)$, $\cQ=16$, $D=0$ model at several couplings}
\label{table5}
\end{center}
\end{table}
Notice that in the case of $\cQ=16$ supercharges
the fractional deviation of the expectation value from its
exact supersymmetric value is smaller for $SU(3)$ than
$SU(2)$ at fixed
lattice coupling which
is consistent with the breaking effect vanishing in the
limit $N\to\infty$. The same
effect was not visible however for $\cQ=4$ supercharges.

Additionally notice that 
table~\ref{table5} contains a fourth column corresponding
to the average value of the cosine of the Pfaffian phase (the average
sine being always consistent with zero). While the Pfaffian for the
$\cQ=16$ supercharge model is generically complex it is possible to
show that it is real positive definite
for the $SU(2)$ model in the matrix model limit. In the case of
$SU(3)$ the Pfaffian in zero dimensions is real 
but not necessarily positive definite
and hence the bosonic action is reweighted with
the sign of the Pfaffian, as discussed
in the previous section. 

\begin{figure}
\begin{center}
\includegraphics[height=55mm]{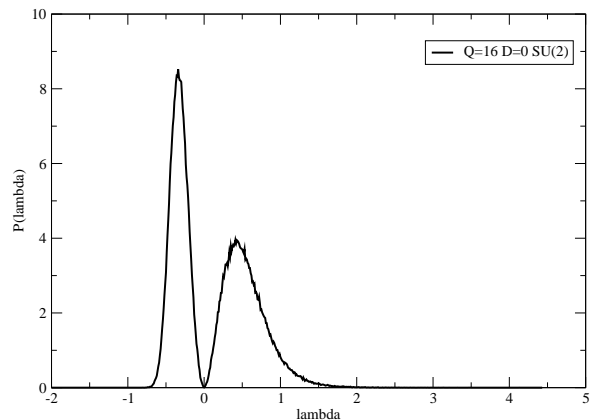}
\caption{Probability distribution of
eigenvalues of $\cU^\dagger_\mu\cU_\mu-I$ for $\cQ=16$, $SU(2)$ and $D=0$}
\label{figure4}
\end{center}
\end{figure}

\begin{figure}
\begin{center}
\includegraphics[height=60mm]{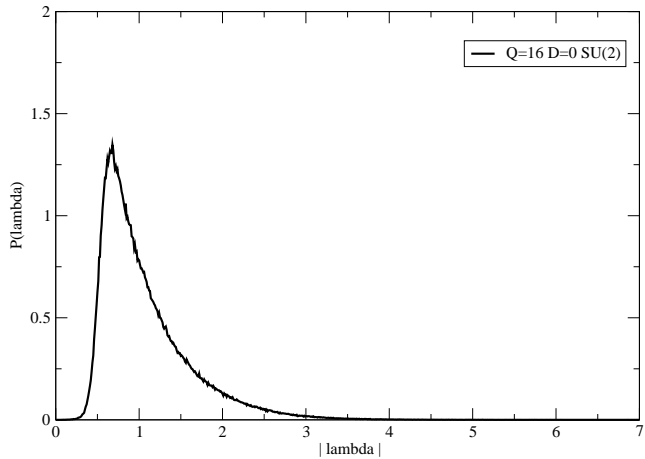}
\caption{Probability distribution of
eigenvalues of the fermion operator for $\cQ=16$, $SU(2)$ and $D=0$}
\label{figure5}
\end{center}
\end{figure}

The probability distribution for $\cU^\dagger_\mu\cU-I$ for
$\cQ=16$ and $SU(2)$ is shown in figure~\ref{figure4}.
While it is qualitatively similar to the $\cQ=4$ case it should be
clear that the localization around the origin is more dramatic than
for $\cQ=4$ -- indeed, theory and simulation point to a larger
value of the power law exponent $p=15$ governing the tail of
the distribution. Since now the scalar fields do not penetrate 
far down the classical flat directions one would expect
to see correspondingly fewer near zero modes in the 
fluctuation spectra of either
the bosonic or, by supersymmetry, fermionic operator in this
case.
Figure~\ref{figure5} shows that indeed this is the case -- a gap appears
to open up in the probability distribution of the magnitude of the
fermion
eigenvalue close to the origin.

We will see that these qualitative features survive in the two
dimensional case to which we now turn.

\section{Two dimensions}

\subsection{$\cQ=4$ supercharges}
We have also begun an investigation of the $\cQ=4$ model with gauge
group $SU(2)$ in two
dimensions. Again, we have focused on the bosonic action, the
distribution of scalar eigenvalues and the spectrum of the fermion
operator. In these simulations we hold fixed the
continuum dimensionless 't Hooft coupling $\lambda \beta^2=0.5$ as
defined in eqn.~\ref{coupling} which leads to a lattice
coupling that grows like the square of the lattice length $L$. 
Table~\ref{table6} shows 
results from simulations on lattices of size $L=2,3,4$ for gauge group
$SU(2)$.
\begin{table}
\begin{center}
\begin{tabular}{||c|c|c|c|c||}\hline
$\kappa$ & $<S_B>^q$ & $<S_B>$ & $S_B^{\rm exact}$& $<\cos{\alpha}>$\\\hline
8.0 &17.15(1)& 17.56(3) & 18.0 & -0.24(1) \\\hline
18.0 & 39.23(2) & 41(6) & $40.5$ & -0.06(2)\\\hline
32.0 & 70.61(4) & 65(5) & 72.0 & -0.014(6) \\\hline
\end{tabular}
\caption{Observables for $SU(2)$ $\cQ=4$ model in $D=2$}
\label{table6}
\end{center}
\end{table}
The second column corresponds to the phase quenched approximation in
which the data is {\it not} reweighted by the Pfaffian phase. The third
column corresponds to the reweighted action. Notice that
the phase quenched
numbers are relatively close to their exact values. However, while there
is a hint from the data that
reweighting pushes them closer to those exact values, it should be
clear that the statistical errors increase rapidly with lattice size
rendering it impossible to make meaningful measurements on the
larger lattices. This is
confirmed by looking at the average of the
cosine of the Pfaffian phase shown in the
fifth column. It is statistically consistent with zero on the larger
lattices. 

\begin{figure}
\begin{center}
\includegraphics[height=60mm]{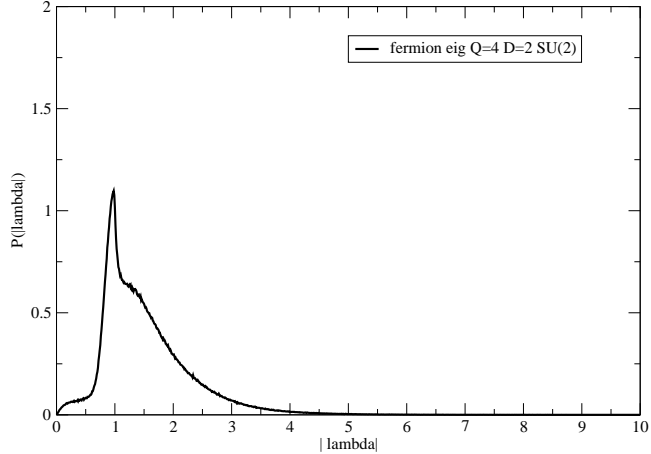}
\caption{Probability distribution of absolute fermion
eigenvalue for $\cQ=4$ $D=2$ $L=2$ and $SU(2)$}
\label{figure6}
\end{center}
\end{figure}

\begin{figure}
\begin{center}
\includegraphics[height=60mm]{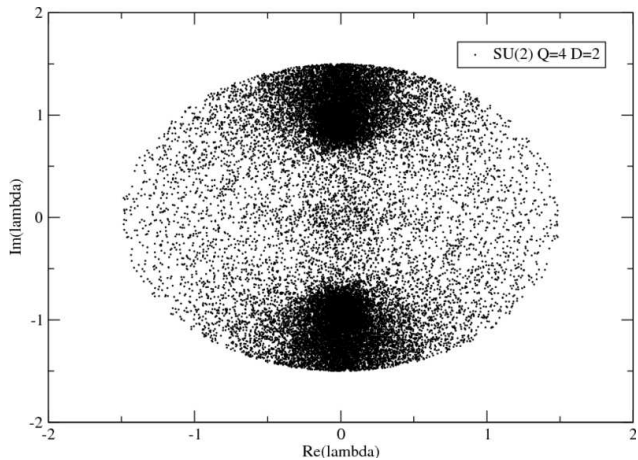}
\caption{
Eigenvalues of the fermion operator for $\cQ=4$, $SU(2)$ and $D=2$}
\label{figure7}
\end{center}
\end{figure}

Actually, the observation that the phase factor
vanishes in this theory is quite interesting; since we
employ periodic boundary conditions the measured
expectation value
yields the Witten index for the theory:
\beq W=<e^{i\alpha}>_{\rm phase\; quenched}\eeq 
Thus our data is consistent with
a vanishing Witten index -- a necessary condition for supersymmetry breaking.
Furthermore, as in the matrix model case, we have observed a
large number of rather small fermion eigenvalues -- figure~\ref{figure6}
shows the distribution of the absolute value of the fermion eigenvalue for
the $SU(2)$ model on a lattice with $L=2$. 
The existence of many small magnitude eigenvalues
is highlighted by the scatter plot of figure~\ref{figure7}
showing the real
and imaginary parts of all fermion eigenvalues with magnitude
$|\lambda|<1.5$ obtained from a sample
of $1000$ configurations for $L=2$. Notice that the eigenvalues, 
while concentrated
in a band along the imaginary axis, nevertheless have a non-zero density, 
by virtue of the Yukawa couplings, over the entire plane
including the region around the origin. As we will
see later this last feature
is {\it not} seen for $\cQ=16$ supercharges where we observe
essentially no eigenvalues in the vicinity of the origin.

For $\cQ=4$ it is possible that these near
massless states could play a role as Goldstino modes associated with dynamical
supersymmetry breaking which 
has been conjectured to happen for the $\cQ=4$ theory in low
dimensions \cite{Hori:2006dk}.

\begin{figure}
\begin{center}
\includegraphics[height=60mm]{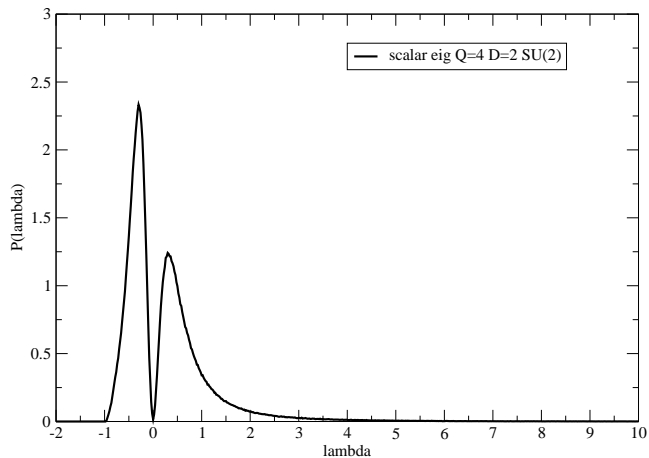}
\caption{Probability distribution of
eigenvalues of $\cU^\dagger_\mu\cU_\mu-I$ for $\cQ=4$, $SU(2)$ and $D=2$}
\label{figure8}
\end{center}
\end{figure}

The scalar eigenvalue distribution is similar to its cousin in zero
dimensions and is shown in figure~\ref{figure8} and shows again that
the scalar distribution possesses a tail to large eigenvalue.
Furthermore, we
have observed that there is a correlation between the
occurrence of small fermion
eigenvalues and the presence of large scalar fields associated
with a significant component of the scalar fields
along the classical flat directions.

Our overall conclusion is that practical simulations of the
$\cQ=4$ supercharge theory in two dimensions with
supersymmetry preserving
periodic boundary conditions will be extremely hard due the rapid
fluctuations in the phase of Pfaffian - a problem which was
first highlighted in \cite{Giedt_det2}.  Furthermore, it is possible that these
phase variations are associated with a dynamical breaking of
supersymmetry as argued for in \cite{Hori:2006dk}. Our numerical
results support a vanishing Witten index and the 
presence of massless fermions -- necessary conditions
for supersymmetry breaking. If so these results contradict
the numerical results presented in \cite{Kanamori:2007yx}. A 
spontaneous breaking
of supersymmetry would also invalidate the theoretical arguments 
given by Matsuura \cite{Matsuura:2007ec} 
showing that the orbifold theories have zero vacuum energy. However
the latter result is derived using 
semiclassical exactness which is invalid if
the $\cQ$-supersymmetry is spontaneously broken by
non-perturbative effects. Clearly, further
work is required to resolve this question unambiguously.

\subsection{$\cQ=16$ supercharges}

The results for the bosonic action for the $\cQ=16$ model are shown in
table~\ref{table7} and correspond to simulations on lattices
of size $L=2,3,4$ and gauge group $SU(2)$.
\begin{table}
\begin{center}
\begin{tabular}{||c|c|c|c|c||}\hline
$\kappa$ & $<S_B>^q$ & $<S_B>$ & $S_B^{\rm exact}$& $<\cos{\alpha}>$\\\hline
8.0 &53.26(6)& 53.26(6)& 54.0 & 0.999997(1)\\\hline
18.0 & 120.1(2)& 120.1(2)& $121.5$ & 0.999995(1)\\\hline
32.0 & 214.7(4) & 214.6(3) & 216.0& 0.999994(3)\\\hline
\end{tabular}
\caption{Observables for $SU(2)$ $\cQ=16$ model in $D=2$}
\label{table7}
\end{center}
\end{table}
Perhaps the most striking result is that the Pfaffian phase is
very close to zero and remains so as the lattice size is
increased and the continuum limit approached. In this case the
reweighting procedure is irrelevant 
as can be seen by comparing the phase quenched and full expectation
values shown in table~\ref{table7}.

\begin{figure}
\vspace{1cm}
\begin{center}
\includegraphics[height=60mm]{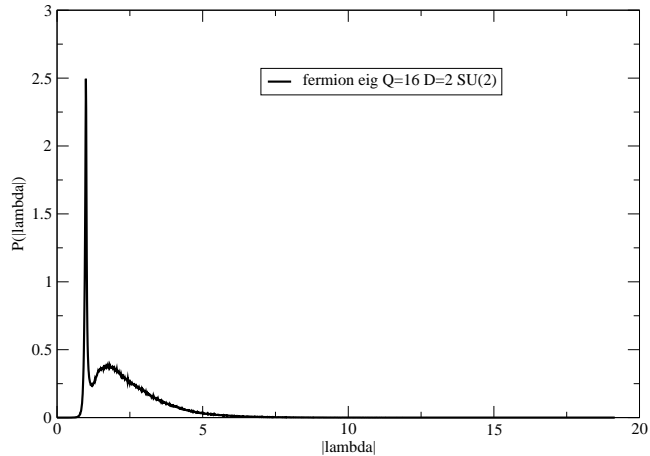}
\caption{Probability distribution of fermion
eigenvalues for $\cQ=16$, $D=2$, $L=2$ and $SU(2)$}
\label{figure9}
\end{center}
\end{figure} 

\begin{figure}
\begin{center}
\includegraphics[height=60mm]{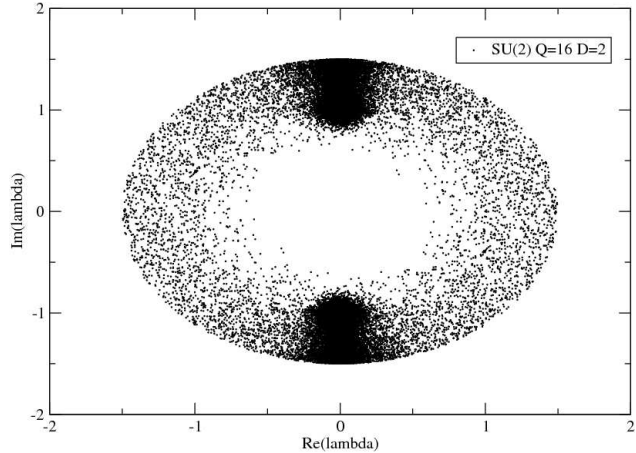}
\caption{
Eigenvalues of the fermion operator for $\cQ=16$, $SU(2)$ and $D=2$}
\label{figure10}
\end{center}
\end{figure}

We believe that this suppression in the Pfaffian phase is related to
the phenomena we observed in zero dimensions; the scarcity of small
fermion eigenvalues. A plot of the distribution of the absolute 
fermion eigenvalue
for this theory is shown in figure~\ref{figure9} and indeed a gap
appears to open up in the spectrum. 
This conclusion is reinforced when we examine a scatter plot of the fermion
eigenvalues in the complex plane given in figure~\ref{figure10}.
Remarkably, and in contrast to the $\cQ=4$ model, the eigenvalues are
excluded from a region around the origin and concentrate along the
imaginary axis.  
It is reasonable to conjecture that
it is the small eigenvalues that control the phase; after all in the limit
in which the eigenvalues are confined to the imaginary axis an eigenvalue
must flow from positive to negative values for a sign change to occur. Such
a transition would require a zero eigenvalue to arise which we
observe to be highly unlikely. Thus phase fluctuations
are suppressed. 

\begin{figure}
\begin{center}
\includegraphics[height=60mm]{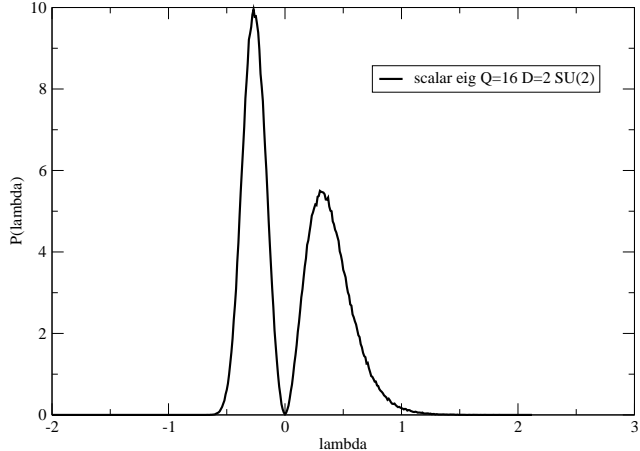}
\caption{Probability distribution of $\cU^\dagger_\mu\cU_\mu-I$
eigenvalues for $\cQ=16$, $D=2$, $L=2$ and $SU(2)$}
\label{figure11}
\end{center}
\end{figure}

Furthermore, the presence of a gap in the fermion
spectrum correlates to a rather rapidly damped scalar
eigenvalue distribution for the $\cQ=16$ model
as we saw in zero dimensions. This distribution is shown in 
figure~\ref{figure11}.
 
\section{Four dimensions}
Finally we report on our preliminary simulations of the $\cQ=16$ supercharge
theory in four dimensions. Table~\ref{table8} shows the bosonic action
and cosine of the Pfaffian phase for lattices with size
$L=2,3$ at fixed
't Hooft coupling $\lambda=0.5$ (the data corresponds to $6000$ and
$1000$ configurations for $L=2$ and $L=3$ respectively)
\begin{table}
\begin{center}
\begin{tabular}{||c|c|c|c|c||}\hline
$L$ & $<S_B>^q$ & $<S_B>$ & $S_B^{\rm exact}$& $<\cos{\alpha}>$\\\hline
2 &211.2(2)& 211.2(2)& 216.0 & 0.9945(3)\\\hline
3 &1072.8(10)& 1075.0(35)& $1093.5$ & 0.955(6)\\\hline
\end{tabular}
\caption{Observables for $SU(2)$ $\cQ=16$ model in $D=4$ at $\lambda=0.5$}
\label{table8}
\end{center}
\end{table}

While supersymmetry breaking effects are visible at $O(1)\%$ the Pfaffian
phase is small and reweighting offers a reliable way to deal with
the fluctuations at least for small lattices\footnote{The larger
error in the reweighted numbers for $L=3$ reflects only the decreased statistics
available in that case where the Pfaffian is only computed 
every tenth measurement}. It remains to be seen 
whether this is true as the lattice size is increased
but we see these results as encouraging (after all
we have lost control of the phase already with small lattices
for the $\cQ=4$ model). 

For comparison, table~\ref{table9} shows the same quantities for
't Hooft coupling $\lambda=0.25$.
\begin{table}
\begin{center}
\begin{tabular}{||c|c|c|c|c||}\hline
$L$ & $<S_B>^q$ & $<S_B>$ & $S_B^{\rm exact}$& $<\cos{\alpha}>$\\\hline
2 &212.5(3)& 211.5(5)& 216.0 & 0.9964(2)\\\hline
3 &1081.9(17)& 1080.5(45)& $1093.5$ & 0.983(2)\\\hline
\end{tabular}
\caption{Observables for $SU(2)$ $\cQ=16$ model in $D=4$ at $\lambda=0.25$}
\label{table9}
\end{center}
\end{table}
Notice that as we approach weak coupling and smaller lattice spacings
the bosonic action moves towards its exact supersymmetric value as
expected. Furthermore, the phase fluctuations also decrease. The question
of whether reweighting will be practical on larger lattices will
hence depend on which trajectory $\lambda_{\rm bare}(L)$ we must follow
in the $(\lambda,L)$ plane to approach the continuum limit. 

\begin{figure}
\begin{center}
\includegraphics[height=60mm]{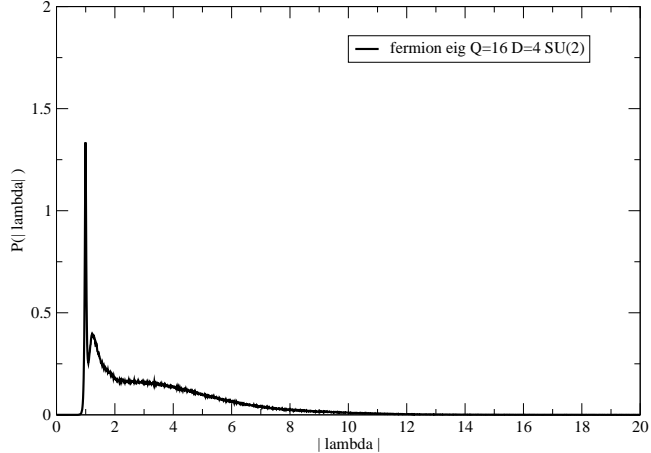}
\caption{Probability distribution of fermion
eigenvalues for $\cQ=16$, $D=4$, $L=2$ and $SU(2)$}
\label{figure12}
\end{center}
\medskip
\end{figure}

\begin{figure}
\begin{center}
\includegraphics[height=60mm]{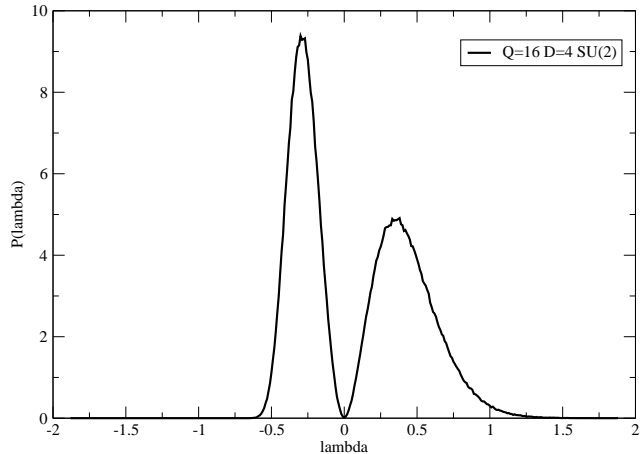}
\caption{Probability distribution of scalar
eigenvalues for $\cQ=16$, $D=4$, $L=2$ and $SU(2)$}
\label{figure13}
\end{center}
\medskip
\end{figure}

The scalar and fermion eigenvalue distributions are shown in
figures~\ref{figure12} and \ref{figure13} and look qualitatively similar to
their two dimensional cousins.

Again, the scatter plot of fermion eigenvalues also indicates that eigenvalues
are repelled from a region around the origin as for the
$\cQ=16$ theory in two dimensions.

\begin{figure}
\begin{center}
\includegraphics[height=60mm]{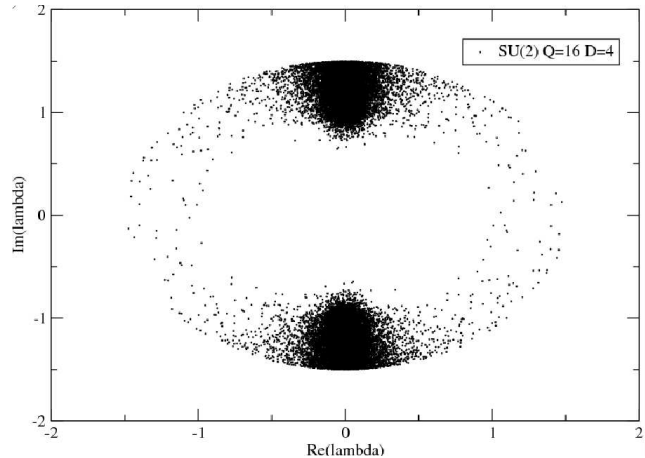}
\caption{
Eigenvalues of the fermion operator for $\cQ=16$, $D=4$ and $SU(2)$}
\label{figure14}
\end{center}
\end{figure}

\section{Discussion}

In this paper we have presented numerical results from simulations of a variety
of Yang-Mills theories with extended supersymmetry. The lattice actions 
that are employed
can be derived either with orbifold methods  
\cite{Cohen:2003xe,Cohen:2003qw,Kaplan:2005ta} or via geometrical
discretization of a twisted version of the target theory 
\cite{Catterall:2007kn}. Remarkably,
they possess both
gauge invariance and one more exact supersymmetries at non-zero lattice spacing.
These lattice theories are important both at a conceptual and practical
level; they offer the possibility of a rigorous definition
of the continuum gauge theory, and through numerical simulation 
may offer up new
ways to extract non-perturbative information on that gauge theory. 
In this paper we have
focused on the latter question; specifically,
are these lattice theories amenable to
Monte Carlo simulation using the tools and techniques of lattice gauge
theory ?

We have shown that the $U(N)$ theories generically suffer from a vacuum
stability problem -- the theory is defined in terms of complexified Wilson
gauge links and these develop one or more zero eigenvalues under 
quantum corrections. This instability is associated with
the trace mode of the would be scalar fields in the lattice theory.
As a result the link fields are driven a long way from the identity
for any value of the lattice spacing rendering invalid the
correspondence between 
the lattice model and the target continuum theory.

However we have shown that this instability of the lattice
action
is avoided if the theory
is truncated to $SU(N)$\footnote{this truncation also has the merit
of removing an exact zero mode of the fermion operator}. 
Furthermore, by measurement of
a simple supersymmetric Ward identity
we have presented evidence that
the associated breaking of supersymmetry is small and decreases as
the continuum limit is taken. 

A second potential problem arises however;
integration over the fermions in these models generically leads to a
{\it complex} Pfaffian (which can be reduced to a determinant for $\cQ=4$
models). Our simulations are necessarily performed in the phase quenched
ensemble where this phase is ignored. In principle, expectation values
can be computed in the full ensemble by a reweighting procedure. However,
this is only practical if the phase fluctuations are small. In the case of
the $\cQ=4$ theory we find the average phase factor approaches zero rapidly as
the continuum limit is taken and the statistical errors in the reweighted
observables grow uncontrollably. This seems to rule out the possibility
of using these actions to study the $\cQ=4$ theory at least for zero
temperature (periodic temporal boundary conditions for all fields).

Intriguingly, the vanishing expectation value of the phase 
factor $<e^{i\alpha}>_{\rm phase\;quenched}=0$
may be given a physical interpretation - it corresponds naively
to a vanishing Witten index for the model. Of course the notion of a Witten
index as a signed sum over classical vacua is
somewhat delicate in theories with
extended supersymmetry which possess a continuum of such vacua. Nevertheless,
a vanishing Witten index is a necessary condition for supersymmetry breaking
and our analysis of the fermion spectrum shows that indeed this theory 
contains one or more
near zero modes which could play the role of a Goldstino associated
with dynamical breaking of supersymmetry. An argument for such a breaking
has been independently by Hori et al \cite{Hori:2006dk} for the
$\cQ=4$ theory in two dimensions. 

Furthermore, we have observed that
the appearance of small fermion eigenvalues is correlated with
large excursions of the scalar fields along the classical directions. In
the background of such a vacuum configuration the operator
describing small fluctuations of the
scalars develops a small eigenvalue corresponding to motion along the
flat direction. This eigenvalue is mirrored in the fermionic sector
because of supersymmetry. The question of whether this effect can lead to 
supersymmetry breaking is then tied to the question of how frequently,
in the context of a Monte Carlo simulation,
the scalars probe these flat directions which, in turn,
is measured by the probability
distribution of scalar field eigenvalues. For $\cQ=4$ we observe that
this distribution $P(\lambda)$ shows a slow power law decay 
which has been estimated to vary as $\lambda^{-3}$ for large $\lambda$
\cite{Krauth:1998xh,Krauth:1999qw}.
This could supply the effect we have argued for; the variance of $\lambda$
then diverges logarithmically and the typical scalar field configuration
wanders appreciably away from the origin and leads to a non-zero density
of small fermion eigenvalues.

At first sight the strong phase
fluctuations visible in the lattice $\cQ=4$ theory are surprising since
a proof exists that the continuum
Pfaffian is real positive semi-definite and one might have therefore
expected the phase fluctuations to be suppressed for small
enough lattice spacing. This does not appear to be
the case. However, we believe there is no inconsistency; if
supersymmetry breaks spontaneously as we claim, the 
Pfaffian will be {\it zero} on the important continuum field
configurations dominating the path
integral, which is then consistent with the vanishing of the expectation
value of the phase factor
seen in the lattice simulations.

In the case of the $\cQ=16$ supercharge model the corresponding scalar
distribution falls much more rapidly with eigenvalue. Correspondingly, we observe that the
fermion spectrum shows a complete absence of near zero modes and the
Pfaffian phase is typically small and can handled, at least for these
small lattices and for $SU(2)$, by reweighting. This is naively
rather unexpected -- afterall the Pfaffian is generically
complex for the $\cQ=16$ model. However, as we
have argued previously,
the average phase factor in the phase quenched
ensemble is nothing more than the
Witten index of the model
which is a topological invariant. It can
hence be evaluated {\it exactly} in the semi-classical limit along the
lines described in \cite{Matsuura:2007ec}
provided supersymmetry remains unbroken which is thought to be the
case for the $\cQ=16$ model. Such a calculation
shows that $W=1$. Thus topological
arguments would suggest that it should be possible to handle
the Pfaffian phase for the $\cQ=16$ theory using reweighting techniques
even though the Pfaffian of the theory is generically complex. This is a remarkable
result but quite consistent with the results of our simulations.

The absence of small fermion eigenvalues together with
the seeming relative unimportance of Pfaffian phase fluctuations
lead us to feel cautiously optimistic about the feasibility of
conducting large scale simulations of the $\cQ=16$ supercharge
theory in four dimensions in the near future. Such simulations
are of course very interesting from the point of view of
the AdSCFT correspondance and we hope to report
on results from such simulations
in the near future \cite{current}.

\acknowledgments Catterall is supported in part by DOE grant
DE-FG02-85ER40237. 

\newpage

\bibliographystyle{JHEP}
\bibliography{orb}

\end{document}